\documentclass[twocolumn]{aastex61}

\newcommand\gs{\mathrel{\raise0.35ex\hbox{$\scriptstyle >$}\kern-0.6em
\lower0.40ex\hbox{{$\scriptstyle \sim$}}}}
\newcommand\ls{\mathrel{\raise0.35ex\hbox{$\scriptstyle <$}\kern-0.6em
\lower0.40ex\hbox{{$\scriptstyle \sim$}}}}

\shorttitle{Counts of Submillimeter Galaxies from the ALMA/SCUBA-2 survey AS2UDS}
\shortauthors{Stach et al.}

\begin{document}

\title{An ALMA survey of the SCUBA-2 Cosmology Legacy Survey UKIDSS/UDS field:
Number counts of submillimeter galaxies}

\correspondingauthor{Stuart M.\ Stach}
\email{stuart.m.stach@durham.ac.uk}

\author[0000-0003-1122-6948]{Stuart M.\ Stach}
\affil{Centre for Extragalactic Astronomy, Department of Physics, Durham University, Durham, DH1 3LE, UK}

\author{Ian Smail}
\affil{Centre for Extragalactic Astronomy, Department of Physics, Durham University, Durham, DH1 3LE, UK}

\author{A.\,M.\ Swinbank}
\affil{Centre for Extragalactic Astronomy, Department of Physics, Durham University, Durham, DH1 3LE, UK}

\author{J.\,M.\ Simpson}
\affil{Academia Sinica Institute of Astronomy and Astrophysics, No. 1, Sec. 4, Roosevelt Rd., Taipei 10617, Taiwan}
\author{J.\,E.\ Geach}
\affil{Centre for Astrophysics Research, School of Physics, Astronomy and Mathematics, University of Hertfordshire, Hatfield AL10 9AB, UK}
\author{Fang~Xia An}
\affil{Purple Mountain Observatory, China Academy of Sciences, 2 West Beijing Road, Nanjing 210008, China}
\affil{Centre for Extragalactic Astronomy, Department of Physics, Durham University, Durham, DH1 3LE, UK}
\author{Omar Almaini}
\affil{School of Physics and Astronomy, University of Nottingham, University Park, Nottingham, NG7 2RD, UK}
\author{Vinodiran Arumugam}
\affil{European Southern Observatory, Karl Schwarzschild Strasse 2, Garching, Germany}
\affil{Institute for Astronomy, University of Edinburgh, Royal Observatory, Blackford Hill, Edinburgh EH9 3HJ, UK}
\author{A.\,W.\ Blain}
\affil{Department of Physics and Astronomy, University of Leicester,University Road, Leicester LE1 7RH, UK}
\author{S.\,C.\ Chapman}
\affil{Department of Physics and Atmospheric Science, Dalhousie University Halifax, NS B3H 3J5, Canada}
\author{Chian-Chou Chen}
\affil{European Southern Observatory, Karl Schwarzschild Strasse 2, Garching, Germany}
\author{C.\,J.\ Conselice}
\affil{School of Physics and Astronomy, University of Nottingham, University Park, Nottingham, NG7 2RD, UK}
\author{E.\,A.\ Cooke}
\affil{Centre for Extragalactic Astronomy, Department of Physics, Durham University, Durham, DH1 3LE, UK}
\author{K.\,E.\,K.\ Coppin}
\affil{Centre for Astrophysics Research, School of Physics, Astronomy and Mathematics, University of Hertfordshire, Hatfield AL10 9AB, UK}
\author{J.\,S.\ Dunlop}
\affil{Institute for Astronomy, University of Edinburgh, Royal Observatory, Blackford Hill, Edinburgh EH9 3HJ, UK}
\author{Duncan Farrah}
\affil{Virginia Polytechnic Institute and State University Department of Physics, MC 0435, 910 Drillfield Drive, Blacksburg, VA 24061, USA}
\author{B.\, Gullberg}
\affil{Centre for Extragalactic Astronomy, Department of Physics, Durham University, Durham, DH1 3LE, UK}
\author{W.\, Hartley}
\affil{Department of Physics and Astronomy, University College London, London, WC1E 6BT, UK}
\author{R.\,J.\ Ivison}
\affil{European Southern Observatory, Karl Schwarzschild Strasse 2, Garching, Germany}
\affil{Institute for Astronomy, University of Edinburgh, Royal Observatory, Blackford Hill, Edinburgh EH9 3HJ, UK}
\author{D.\,T.\ Maltby}
\affil{School of Physics and Astronomy, University of Nottingham, University Park, Nottingham, NG7 2RD, UK}
\author{M.\,J.\ Micha\l{}owski}
\affil{Astronomical Observatory Institute, Faculty of Physics, Adam Mickiewicz University, ul. S\l{}oneczna 36, 60-286 Pozna\'n, Poland}
\author{Douglas Scott}
\affil{Department of Physics and Astronomy, University of British Columbia, 6224 Agricultural Road, Vancouver, BC V6T 1Z1, Canada}
\author{Chris Simpson}
\affil{Gemini Observatory, Northern Operations Center, 670 N. A'ohuku Place, Hilo, HI 96720, USA}
\author{A.\,P.\ Thomson}
\affil{The University of Manchester, Oxford Road, Manchester, M13 9PL, UK}
\author{J.\,L.\ Wardlow}
\affil{Centre for Extragalactic Astronomy, Department of Physics, Durham University, Durham, DH1 3LE, UK}
\author{P.\ van der Werf}
\affil{Leiden Observatory, Leiden University, P.O. box 9513, NL-2300 RA Leiden, The Netherlands}

\begin{abstract}
We report the first results of AS2UDS: an 870\,$\mu$m continuum survey with the Atacama Large Millimeter/Submillimeter Array (ALMA) of  a total area of $\sim$\,50\,arcmin$^2$ comprising a complete sample of 716 submillimeter sources drawn from the SCUBA-2 Cosmology Legacy Survey (S2CLS) map of the UKIDSS/UDS field. The S2CLS parent sample covers a 0.96\,degree$^2$ field at $\sigma_{850}=0.90\pm0.05$\,mJy\,beam$^{-1}$. Our deep, high-resolution ALMA observations with $\sigma_{\rm 870}\sim$\,0.25\,mJy and a 0$\farcs$15--0$\farcs$30 FWHM synthesized beam, provide precise locations for 695 submillimeter galaxies (SMGs) responsible for the submillimeter emission corresponding to 606 sources in the low resolution, single-dish map. We measure the number counts of SMGs brighter than $S_{\rm 870}\geq$\,4\,mJy, free from the effects of blending and show that the normalisation of the counts falls by 28\,$\pm$\,2\,\% in comparison to the SCUBA-2 parent sample, but that the shape remains unchanged.  We determine that 44$^{+16}_{-14}$\,\% of the brighter single-dish sources with $S_{\rm 850}\geq$\,9\,mJy  consist of a blend of two or more ALMA-detectable SMGs brighter than $S_{\rm 870}\sim$\,1\,mJy (corresponding to a galaxy with a total-infrared luminosity of $L_{\rm IR}\gs$\,10$^{12}$\,L$_\odot$), in comparison to 28\,$\pm$\,2\,\% for the single-dish sources at $S_{\rm 850}\geq$\,5\,mJy. Using the 46 single-dish submillimeter sources that contain two or more ALMA-detected SMGs with photometric redshifts, we show that there is a significant statistical excess of pairs of SMGs with similar redshifts ($<$\,1\,\% probability of occurring by chance), suggesting that \textit{at least} 30\,\% of these blends arise from physically associated pairs of SMGs. 
\end{abstract}
\keywords{galaxies: starburst -- galaxies: high-redshift}

%
%
%
\section{Introduction} \label{sec:intro}
It has been two decades since the Submillimeter Common User Bolometer Array (SCUBA) instrument on the James Clerk Maxwell Telescope (JCMT) enabled deep observations of high-redshift submillimeter sources which expanded the number of known high-redshift submillimeter luminous infrared sources up to hundreds \citep[e.g.][]{smail1997deep,hughes1998high,barger1998submillimetre}. These submillimeter galaxies (SMGs) constitute a population of the most intensely star-forming galaxies, with star-formation rates (SFRs) in the 100s--1000s of M$_{\odot}$\,yr$^{-1}$  \citep{blain2002submillimeter,magnelli2012herschel,swinbank2013alma,casey2013characterization} at typical redshifts $z\sim $\,2--3 \citep{chapman2005redshift,wardlow2011laboca, simpson2014alma,chen2016scuba}.

This level of star formation means that in a single starburst event, an SMG would need just a few hundred million years to form the stellar mass of a massive galaxy ($M_{\ast} \gtrsim $\,10$^{11}$\,M$_{\odot}$). This has led to the suggestion that SMGs have many of the properties expected for the progenitors of the luminous massive elliptical and spheroid galaxies in the local Universe \citep{lilly1999canada,fu2013rapid,simpson2014alma} with speculation that they could represent a phase in a single evolutionary path linking SMGs to luminous quasi-stellar objects (QSOs) at $z\sim$\,2  and massive, passive galaxies found at $z\sim$\,1--2 \citep{coppin2008testing,cimatti2008gmass,whitaker2012large,toft2014submillimeter}. Further evidence for this evolutionary path comes from clustering studies from single-dish detections, suggesting they reside in halos of mass $\sim $\,10$^{13}$\,M$_{\odot}$, consistent with that of $z\sim$\,2 QSOs and with their subsequent evolution into local ellipticals \citep{farrah2006spatial,hickox2012laboca,wilkinson2016scuba}.

However, whilst SMGs may play a significant role in the stellar mass growth of massive galaxies, measuring their basic properties have been hampered by the coarse angular resolution of the single-dish telescopes, with beams of $\sim$\,15$\arcsec$\,FWHM. One of the questions raised is whether the (coarse resolution) single-dish detections arises from a single SMG or are blends of multiple SMGs within the single-dish beam. To measure the blending and to accurately identify SMG counterparts at other wavelengths requires high-resolution interferometric studies, which were initially performed via radio counterpart identification \citep[e.g.][]{chapman2005redshift,ivison2007scuba}, but more recently with submillimeter interferometers. \cite{wang2010sma} use deep 850\,$\mu$m integrations of two bright submillimeter sources in the GOODS-N field to suggest that both sources break into multiple components and suggested that around 30\% of 850-$\mu$m sources with flux densities ($S_{\rm 850}$) $S_{\rm 850}\geq$\,5\,mJy could be composed of blends of more than one SMG. ALMA observations of much larger samples suggested that this rises to $>$\,90\% for $S_{\rm 850}\sim$\,8\,mJy sources selected in single-dish surveys \citep[e.g.][]{simpson2015scuba2}. More recently, \cite{hill2018high} used the Submillimeter Array (SMA) to observe 75 of the brightest S2CLS sources (S$_{\rm 850}\gtrsim$\,8\,mJy) at 870\,$\mu$m with a resolution of $\sim$\,2$\farcs$4. Combining their SMA data with archival observations they determine a lower multiplicity rate of $\sim$15\,\%, which is consistent with previous work  with the SMA \citep{chen2013resolving}. However these SMA observations are limited by the sensitivity, with \cite{hill2018high} using maps with an average rms depth of $\sim$\,1.5\,mJy. This meant that multiples can only be identified in a bright single-dish source if both components have near equal flux density, which is unlikely to be a frequent occurrence. Therefore, care needs to be taken when comparing such multiplicity studies since they can use different criteria for the brightness ratio  of detected sources.

To make definitive progress in understanding the properties of SMGs area requires the improvements in sensitivity and resolution provided by the Atacama Large Millimeter/Submillimeter Array (ALMA). The first such study, comprising Cycle 0 observations of the 122 submillimeter sources detected in the LABOCA survey of the Extended \textit{Chandra} Deep Field South \citep[LESS:][]{weiss2009large} found that 30\,\% of LABOCA sources resolved into multiple components with $S_{\rm 850}\gtrsim$\,1.5\,mJy when observed at 1$\farcs$5 resolution \citep{karim2013alma,hodge2013alma}. Following this result, in ALMA Cycle 1, 30 of the brightest submillimeter sources (median single-dish flux density of $S_{\rm 850}\gs$\,9\,mJy) from the SCUBA-2 Cosmology Legacy Survey \citep[S2CLS:][]{geach2017scuba} map of the UKIDSS Ultra Deep Survey (UDS, \citealt{lawrence2007ukirt}) field were observed with ALMA by \citet{simpson2015scuba2}. This confirmed that the majority (61$^{+19}_{-15}$\,\%) of bright, single-dish  submillimeter sources are comprised of blends of multiple SMGs brighter than  $S_{\rm 850}\sim$\,1.5\,mJy \citep{simpson2015scuba2,simpson2015scuba}. Each of these bright single-dish sources consists of 2--4 SMGs, which themselves are ultraluminous infrared galaxies (ULIRGs; $L_{\rm IR}\geq10^{12}L_{\odot}$), seen within a projected diameter of  $\sim$\,150\,kpc. \cite{simpson2015scuba2} suggest that such a high over-density of SMGs requires that the majority of such detections result from physical association, as opposed to chance projections along the line of sight. 

Several studies have used spectroscopic observations of molecular gas emission to test the origin of blends of SMGs.  For example, \cite{zavala2015early} used spectroscopic detections for the components in one blended submillimeter-bright lensed galaxy to show that it split into three distinct galaxies, each at significantly different redshifts.   More recently,  \cite{wardlow2018} used ALMA observations to search for CO emission in the fields of six submillimeter sources, which include a total of 14 SMGs, to determine that $\gtrsim$\,75\% of blends of multiple SMGs are not physically associated.  Similarly,
\cite{hayward2018observational} report optical and near-infrared spectroscopy of a sample of seven single-dish sources, where three showed a blending of physically associated SMGs, whilst four contained at least one pair of components that was physically unassociated. This mix  of physically associated and unassociated components in the blended single-dish submillimeter sources is consistent with semi-analytic modelling, for example \cite{cowley2014simulated} have suggested that most blends of SMGs in single-dish sources arise from projections of unrelated galaxies seen along the line of sight.

The presence of multiple SMG counterparts to individual single-dish submillimeter sources  indicates that the number counts derived from low-resolution single-dish surveys do not represent the true number counts of SMGs.  Even a small change in the expected form of the counts of SMGs has a potentially significant impact on models that use them as a constraint on the evolution of high-redshift, dust obscured starbursts \citep[e.g.][]{cowley2014simulated,lacey2016unified}.  

In this paper we present the first results of the recently completed ALMA survey of the full S2CLS UDS sample, which comprises 870\,$\mu$m maps of the 716 $>$\,4\,$\sigma$ single-dish sources with observed $S_{\rm 850}\geq$\,3.4\,mJy  in this 0.96\,degree$^2$ field. Our deep, high-resolution ALMA survey, with rms depths of $\sigma_{\rm 870}\sim$\,0.25\,mJy\,beam$^{-1}$ at 0$\farcs$15--0$\farcs$30 resolution, provides the statistical sample necessary to study the SMG population in detail and supplies us with the largest sample of ALMA-detected SMGs currently available. From this we construct resolved  870-$\mu$m SMG number counts and investigate the multiplicity in single-dish surveys. In \S \ref{sec:data} we describe the sample selection, observations, data reduction and source extraction. \S \ref{sec:results} covers our results and discussions and \S \ref{sec:conclusion} gives our conclusions.

%
%
%
\section{Observations and Data Reduction} \label{sec:data}

\subsection{Sample Selection}

Our survey (the ALMA-SCUBA-2 Ultra Deep Survey field survey, hereafter AS2UDS) is based on a complete sample of 850-$\mu$m sources selected from the S2CLS map of the UDS field \citep{geach2017scuba}. The S2CLS UDS map covers an area of 0.96\,deg$^{2}$, with noise levels below 1.3\,mJy and a median depth of $\sigma_{\rm 850}=$\,0.88\,mJy\,beam$^{-1}$ with 80\% of sources having $\sigma_{\rm 850}=$\,0.86--1.02\,mJy\,beam$^{-1}$. Between Cycles 1, 3 and 4 we observed all 716 $>4\sigma$ sources from the SCUBA-2 map,  giving an observed flux density limit of $S_{\rm 850}\geq $\,3.4\,mJy, or a deboosted flux density of $S^{\rm deb}_{\rm 850}\geq$\,2.5\,mJy
\citep{geach2017scuba}.  

As a pilot project in Cycle 1 (Project ID: 2012.1.00090.S), 30 of the brightest sources from an early version of the SCUBA-2 map (data taken before 2013 February) were observed in Band 7 \citep{simpson2015scuba,simpson2015scuba2,simpson2017scuba}. This early version of the map had a depth of  $\sigma_{\rm 850}\sim$\,2.0\,mJy\,$^{-1}$ and subsequent integration time scattered three of these sources below our final sample selection criteria, leaving 27 of these original single-dish detected sources in our final sample. In Cycles 3 and 4 (Project ID: 2015.1.01528.S and 2016.1.00434.S, respectively) we observed the remaining 689 single-dish sources in the final S2CLS catalog. To cross calibrate the data, a fraction of these sources were observed twice in Cycles 3 and 4 or twice in Cycle 4.
 
\subsection{Data Reduction and Source Detection} 
Full details of the data reduction and source detection will be presented in Stach et al.\ (in prep.) but here we provide a brief overview. Our ALMA targets were observed in Band 7 (344\,GHz\,$\sim$\,870\,$\mu$m), where the frequency closely matches the central frequency of the SCUBA-2 filter transmission and the FWHM of the ALMA primary beam at this frequency (17$\farcs$3) comfortably covers the whole of the SCUBA-2 beam (14$\farcs$7 FWHM). Cycle 1 observations were carried out on 2013 November 1, Cycle 3 between 2016 July 23 and August 11 and Cycle 4 between 2016 November 9 and 17 and 2017 May 6. 

The phase center for each pointing was set to the SCUBA-2 positions from the S2CLS DR1 submillimeter source catalog \citep{geach2017scuba}, with observations taken with 7.5\,GHz bandwidth centred at 344\,GHz using a single continuum correlator set-up with four basebands. Observations of 40\,seconds were employed with the aim to yield 0$\farcs$3 resolution maps with a depth of $\sigma_{\rm 870}=$\,0.25\,mJy\,beam$^{-1}$. However, the Cycle 3 observations were taken in a more extended ALMA configuration, yielding a median synthesised beam of 0$\farcs$19 FWHM.

Calibration and imaging were carried out with the \textsc{Common Astronomy Software Application} \citep[\textsc{casa} v4.6.0;][]{mcmullin2007casa}. For source detection we created ``detection'' maps by applying a 0$\farcs$5 FWHM Gaussian taper in the \textit{uv}-plane, to ensure sensitivity to extended flux from our SMGs that might fall below our detection threshold, as well as  improving efficiency for selecting extended sources. This down-weighting of the long baseline information results in final ``detection'' maps with a mean synthesized beam size of 0$\farcs$73\,$\times$\,0$\farcs$59 for Cycle 1, 0$\farcs$56\,$\times$\,0$\farcs$50 for Cycle 3 and 0$\farcs$58\,$\times$\,0$\farcs$55 for Cycle 4.

The \textsc{clean} algorithm was used to create the continuum maps using multi-frequency synthesis mode with a natural weighting to maximise sensitivity. We initially created a dirty image from the combined spectral windows (SPWs) for each field and calculated the rms noise values. The fields were then initially cleaned to 3\,$\sigma$ and then masking ellipses are placed on sources above 4\,$\sigma$ and the sources are then cleaned to 1.5',$\sigma$. The final cleaned, \textit{uv}-tapered detection maps have mean depths of $\sigma_{\rm 870}=$\,0.25\,mJy\,beam$^{-1}$ for Cycle 1, $\sigma_{\rm 870}=$\,0.34\,mJy\,beam$^{-1}$ for Cycle 3 and $\sigma_{\rm 870}=$\,0.23\,mJy\,beam$^{-1}$ in Cycle 4, the differences here largely being due to the varying resolutions of the observations in each ALMA cycle.

For source detection, \textsc{sextractor} was initially used to find $>$\,2\,$\sigma$ peaks within the ``detection'' maps. Noise estimates were then calculated from the standard deviation in the integrated fluxes in 100 randomly placed 0$\farcs$5 diameter apertures in each map. These were then used, along with the 0$\farcs$5 diameter flux measured for each
detection, to determine the signal-to-noise ratio (SNR) of the sources. As we used an aperture smaller than the beam size the mean 0$\farcs$5 aperture depths in the detection maps are approximately a factor of two deeper than the noise per beams quoted above (with the caveat of a corresponding aperture correction). 

\begin{figure}[ht!]
\plotone{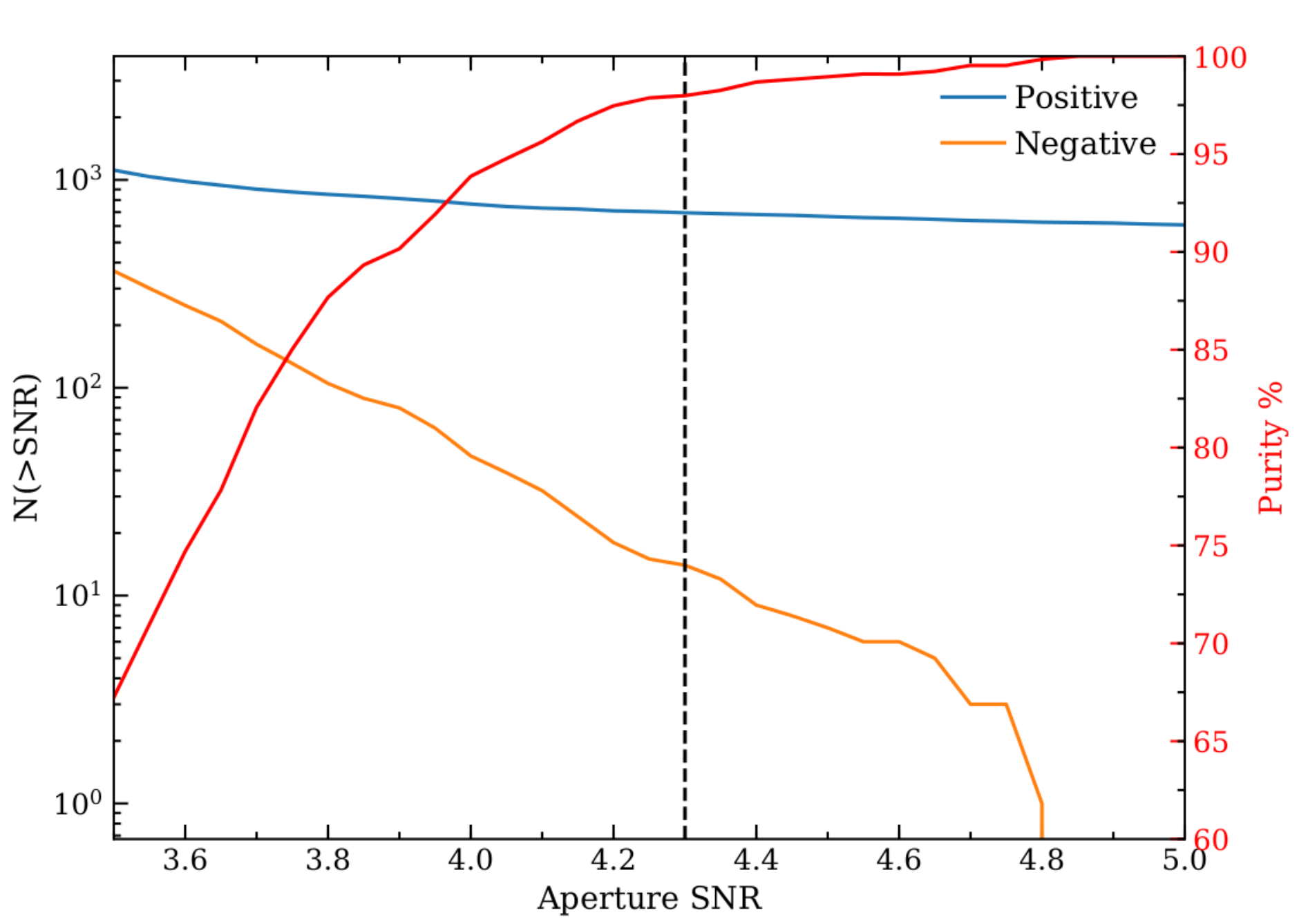}
\caption{The cumulative numbers of sources detected in our 716 ALMA maps above a given signal-to-noise ratio  in both the tapered detection maps (Positive) and the inverted detection maps (Negative).  We select a  SNR threshold for the final AS2UDS catalog which  minimises the contamination from spurious detections, as estimated from the number of equivalent SNR negative sources. We show the corresponding ``purity'' as a function of  SNR threshold and mark our adopted 4.3\,$\sigma$ threshold (dashed line), which yields a 98\,\% purity, equivalent to 14 false sources  in a final catalog of  695 SMGs. \label{fig:purity}
}
\end{figure}

The choice of the  size of the detection aperture and the SNR cut for the sample selection were made based on a trade-off between
purity and depth of the catalog.  
The final catalog consists of the 695\footnote{We detect the strongly lensed SMG 'Orochi' \citep{ikarashi2011detection} but remove this from our analysis.} sources that have a 0$\farcs$5 aperture SNR  $\geq$\,4.3 and fall within the primary beam of the ALMA maps. This threshold and aperture size was chosen to give us a 98\,\% purity rate, $P_{\rm r}$ (2\,\% contamination), calculated as follows:

\begin{equation}
  P_{\rm r}=\frac{N_{p}-N_{n}}{N_{p}},
	\label{eq:doublepower}
\end{equation}
where $N_{p}$ is the number of positive sources detected above the chosen SNR limit (i.e.\ 695) and $N_{n}$ is the number of sources detected above the same limit in the inverted detection maps (made by multiplying the detection maps by $-1$, Figure~\ref{fig:purity}).

We confirm the behaviour of the noise in our maps by comparing our number of ``negative'' sources from the inverted maps at our selected SNR threshold against that expected from a simple Gaussian distribution of independent synthesised beams  \citep{dunlop2016deep}. In AS2UDS, for our average restored beam size, there are roughly $\sim$\,450,000 independent beams across the 716 ALMA pointings. For Gaussian statistics we would then expect $\sim$\,8 ``negative'' sources at 4.3\,$\sigma$. However, as noted by \cite{dunlop2016deep}, based on \cite{condon1997errors,condon1998nrao}, there are effectively twice as many statistically independent noise samples as one would expect from a naive Gaussian approach due to the non-independence of pixel values in synthesised imaging. This would result in an expected $\sim$\,16 ``negative'' sources or 2.3\,$\pm$\,0.5\,\%, which is consistent with the number we detect.

%
%

For each of the detected sources we then derived a 1$\farcs$0 diameter aperture flux density from the primary beam corrected maps, these flux densities are aperture corrected and flux deboosted using the same methodology as \cite{simpson2015scuba2}, as briefly described below.

\subsection{Completeness \& Flux Deboosting}
To calculate the completeness  and flux deboosting factors for our ALMA catalog we inserted model sources into simulated ALMA maps and determined the properties of those which were recovered.   We start with simulated noise maps, to make these as realistic as possible we used ten residual maps output from {\sc casa} (i.e.\ an observed ALMA map where the source flux from any detected sources has been removed). The maps were  selected  to match the distribution in observed $\sigma_{\rm 870}$ for all 716 AS2UDS pointings. Model sources with flux densities drawn from a steeply declining power-law distribution with an index of $-2$, consistent with \cite{karim2013alma,simpson2015scuba2}, and intrinsic FWHM sizes drawn uniformly from a range 0--0$\farcs$9, were convolved with ALMA synthesized beams and inserted into 60,000 simulated noise maps. Then we applied our source detection algorithm and measured recovered fluxes as detailed above, with a successful recovery claimed for detections within the size of a synthesized beam, i.e.\ 0$\farcs$6, from the injected model source position.

The result of these simulations is that we estimate our catalog is 98\,$\pm$\,1\,\% complete for all our simulated sources at $S_{870}\geq$\,4\,mJy, with the incompleteness exclusively arising from the most extended simulated sources (intrinsic FWHM $>$\,0$\farcs$6). As found in \cite{franco2018goods} our simulated maps show the intrinsic sizes of the submillimeter galaxies strongly effects the completeness fractions at low signal-to-noise.  But, at our 4\,mJy threshold we are only miss  a small number of the most extended galaxies.  We note that our simulated sources had sizes which were uniformly distributed up to 0$\farcs$9, whereas previous studies suggest median submillimeter sizes of $\sim0\farcs3$ \citep{tacconi2006high,simpson2015scuba} therefore the 98$\pm$1\,\% completeness is probably conservative.

We estimate the   flux boosting, the effect of noise fluctuations in the overestimation of a source's flux density, by calculating the ratio of the flux density for each recovered  simulated source  to the original input flux density. The fact that noise in the maps is approximately Gaussian, combined with the steep counts of faint sources, means that we find that fluxes are typically overestimated in the lower flux bins. However, again brighter than $S_{870}\geq$\,4\,mJy the flux deboosting becomes a minor correction with a median correction factor of 0.98\,$\pm$\,0.04 for the SMGs considered in this paper.

The complete catalog of  SMGs from AS2UDS, with full descriptions of the source extraction, flux density measurements and flux deboosting will be presented in Stach et al.\ (in prep.).

%
%
%
\section{Analysis, Results and Discussion} \label{sec:results}
The AS2UDS catalog contains 695 SMGs  (detected in 606 ALMA maps), with $S_{\rm 870}\geq$\,0.9\,mJy (4.3\,$\sigma$), across 716 ALMA fields centred on $>4\sigma$ single-dish submillimeter  sources from S2CLS  \citep{geach2017scuba}. The total area of the primary-beam coverage in our ALMA survey is equivalent to 47.3\,arcmin$^2$.  

The AS2UDS SMG sample is roughly seven times larger than the previous largest sub/millimeter interferometric survey of single-dish submillimeter sources \citep[ALESS:][]{hodge2013alma,karim2013alma} and drawn from a field which is four times larger in terms of contiguous area. As was also found in ALESS, a  fraction of our ALMA maps do not contain any detected SMGs (above 4.3\,$\sigma$ significance) -- there are 108 of these ``blank'' maps (15\,$\pm$\,2\,\% of the survey).   In addition, we have 79  maps (11\,$\pm$\,1\,\%) where the single-dish SCUBA-2 source breaks up into multiple SMGs at ALMA resolution. In \S3.2 we show that the blank maps may in part be a result of similar ``multiplicity'' effects, as opposed to false positive detections in the original SCUBA-2 catalog.

With this nearly order-of-magnitude increase in the sample of SMGs, in this paper we present number counts of  SMGs brighter than $S_{\rm 870}\sim$\,4\,mJy, above the original 4-$\sigma$ limit of the single-dish SCUBA-2 survey.  We also utilise the available multi-wavelength data for the UKIDSS/UDS field to employ photometric redshifts for our SMGs to quantify what fraction of the SCUBA-2 sources corresponding to multiple ALMA SMGs are due to chance projections, rather than physical associations.

\subsection{Flux Recovery}
We start by determining the fraction of the original SCUBA-2 sources fluxes which are recovered in the sources we detect in the corresponding maps from ALMA.  In the flux regime that we are interested in  this paper, $S_{\rm 870}\geq$\,4\,mJy, we find that we recover a median fraction of 97$^{+1}_{-2}$\,\% of the original SCUBA-2 flux from SMGs detected within the ALMA primary beam pointing of the corresponding SCUBA-2 parent source.   

In respect of the ``blank'' maps: both the noise properties of the SCUBA-2 sources which resulted in``blank'' maps and the noise properties of the ALMA observations of these maps are indistinguishable from those where ALMA detected an SMG. This suggests that these ``blank'' maps are not simply due to variations in the quality of the input catalog or follow-up observations.  
Similarly, it could be that many of the ``blank'' map sources are due to spurious false positives in the S2CLS parent sample.  We test this by stacking \textit{Herschel}/SPIRE maps at the locations of the 108 ``blank'' map sources, ranked in five bins of their SCUBA-2 flux. We recover emission in all the SPIRE bands (250, 350 and 500 $\mu$m) with flux densities between 7--20\,mJy for all five flux bins. Even for the faintest 10\,\% of SCUBA-2 sources with corresponding ``blank'' ALMA maps,  we still recover SPIRE detections at 250 and 350\,$\mu$m.  Hence we are confident that the majority of the ``blank'' maps are a result of genuine non-detections in ALMA and not false positive sources in the S2CLS map.  However, these  ``blank'' maps do typically correspond to   fainter single-dish sources: the median flux of the ``blank'' maps is $S_{\rm 850}=$\,4.0\,$\pm$\,0.1\,mJy, compared to $S_{\rm 850}=$\,4.5\,$\pm$\,0.1\,mJy for the whole sample. Thus it is possible that a strong increase in flux boosting in the original S2CLS catalog at SNR of $\lesssim$\,4--4.5\,$\sigma$ ($S_{\rm 870}\sim$\,3.6--4.0\,mJy) may play a part in explaining why ALMA detects no SMGs in these maps.  To remove this concern, in our analysis  we only consider the number counts brighter than $S_{\rm 870}\geq$\,4\,mJy. 

We conclude that with the sensitivities of our ALMA maps we can detect $S_{\rm 870}=$\,4\,mJy SMGs in even the shallowest AS2UDS maps across the entirety of the primary beam. In addition, based on our simulated ALMA maps described above, we have shown we have with reliable measured flux densities for the complete sample of 299 $S_{\rm 870}\geq$\,4\,mJy SMGs in the AS2UDS catalog presented here.

\subsection{Number Counts}
\begin{figure*}
\plottwo{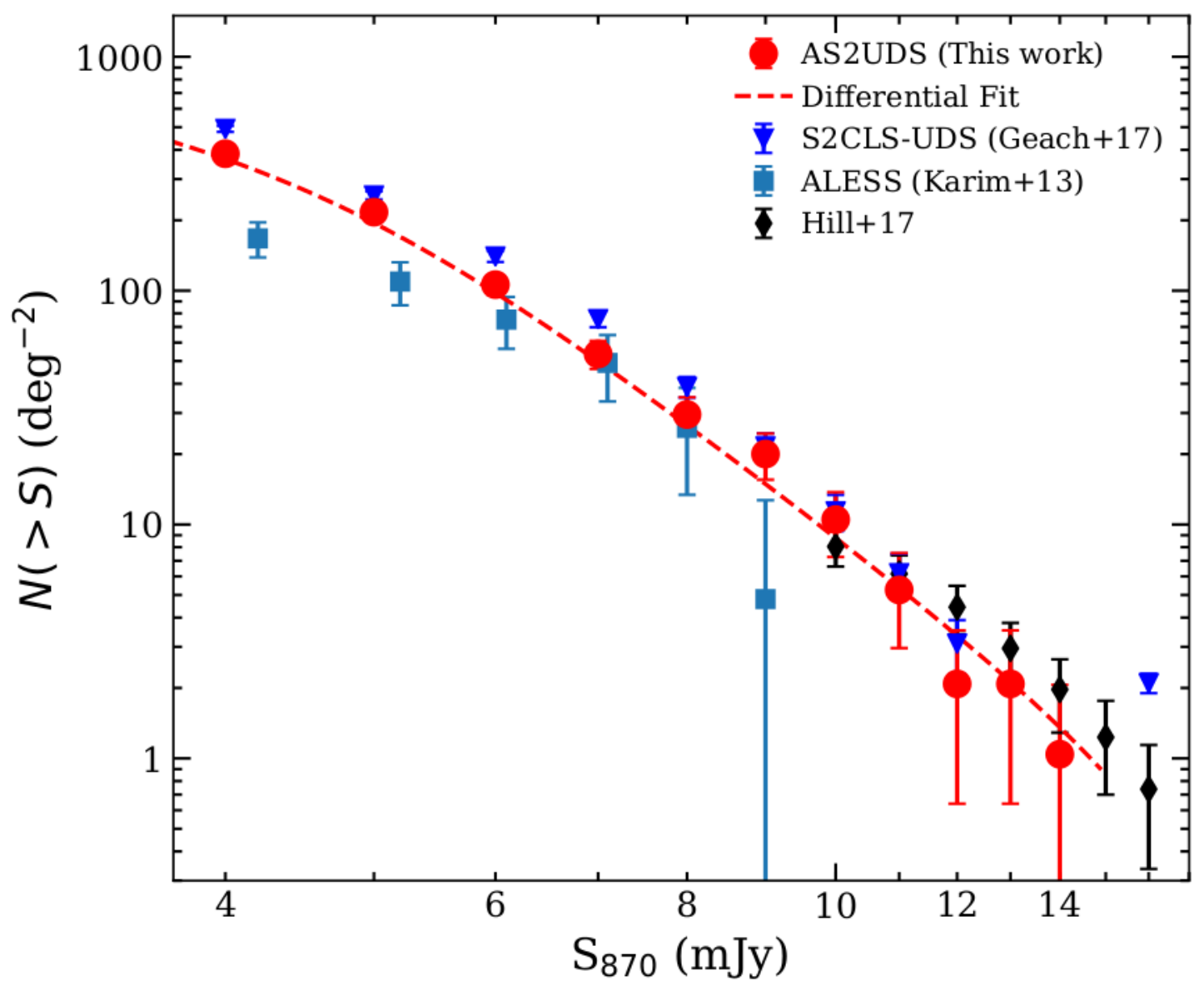}{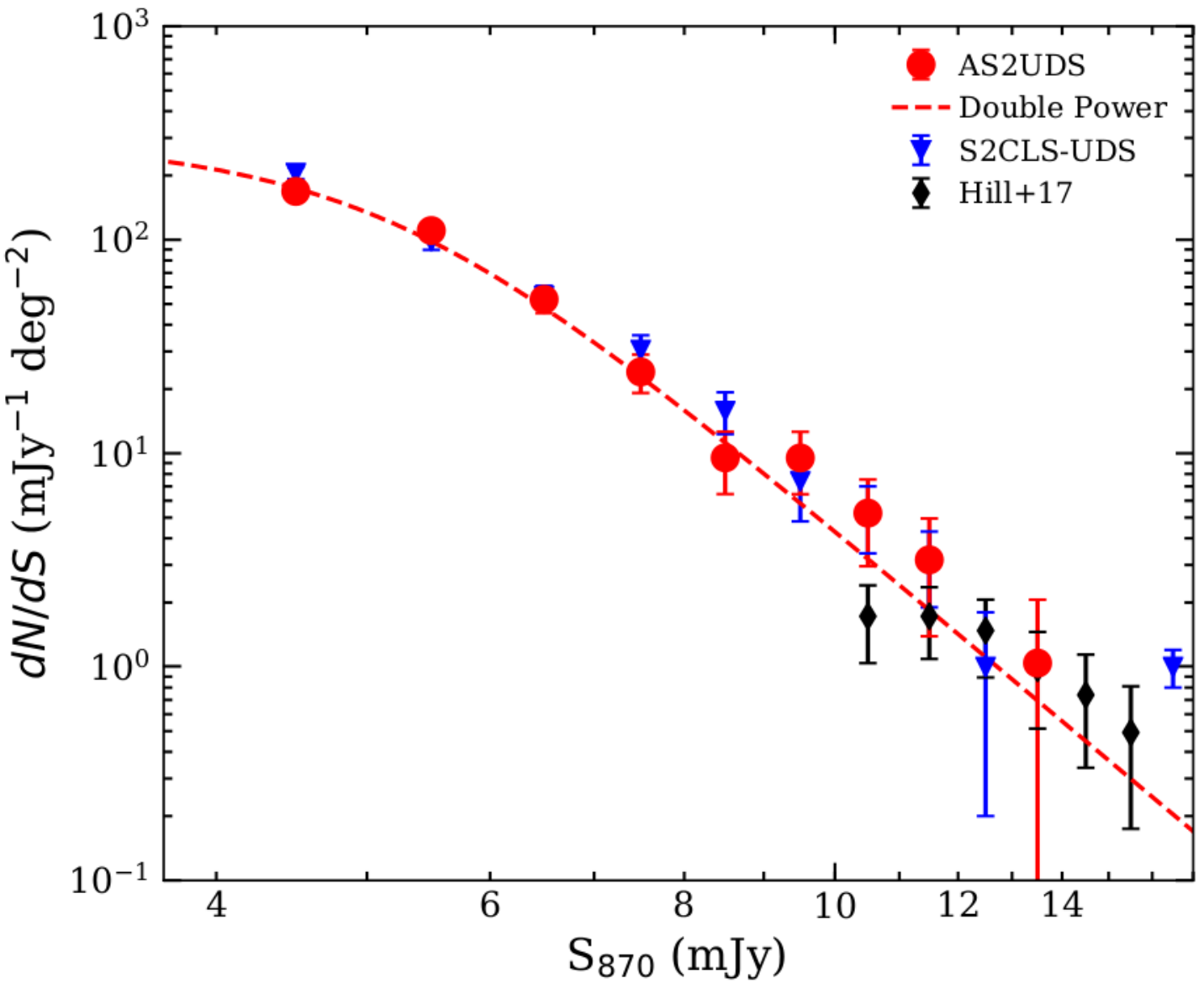}
\caption{\textit{Left:} The 870\,$\mu$m cumulative number counts of ALMA-identified SMGs from the AS2UDS survey. For comparison we also show the original (deboosted) S2CLS single-dish counts for this field \citep{geach2017scuba}, the earlier interferometric SMG counts from ALESS survey \citep{karim2013alma}, as well as those derived from SMA follow-up counts of the brightest S2CLS sources from \citep{hill2018high}. The AS2UDS counts roughly follow the same shape as the parent single-dish counts from S2CLS, but there is a systematic reduction in the surface density of SMGs of the order 37\,$\pm$\,3\,\% (see \S 3.1). The  dashed line is the integral of the double-power law fit to the differential number counts. \textit{Right:} The 870-$\mu$m differential number counts for AS2UDS compared to the parent S2CLS-UDS. A double-power law functional fit is overlaid as a dashed line, and the fitting parameters are given in \S 3.2.}
\label{fig:Counts}
\end{figure*}

In Figure~\ref{fig:Counts}, we show the cumulative and differential number counts of the 299  870\,$\mu$m-selected SMGs from AS2UDS to a flux limit of $S_{\rm 870}=4$\,mJy. Both the cumulative and differential number counts are normalized by the area of the S2CLS UDS map from which the original targets were selected: 0.96\,degree$^{2}$. Whilst the ALMA completeness factors are minimal for AS2UDS the number counts do have to be adjusted for the incompleteness of the parent S2CLS survey. We correct our counts by factoring in the estimated incompleteness of the catalog of the S2CLS UDS map from \cite{geach2017scuba} who reported that the parent sample is effectively complete at $\geq$\,5\,mJy, dropping to  $\sim$\,88\% at $\geq$\,4.5\,mJy and  $\sim$\,83\%  at  $\geq$\,4\,mJy.

As in \cite{karim2013alma} the errors are calculated from both the Poissonian error and the individual flux uncertainties added in quadrature, where the flux uncertainty error is the standard deviation of the mean of the counts for each bin based on 1,000 re-samples of the catalog, assigning random flux densities to each source within their individual error margins, Table 1. We also compare these counts to those from the parent single-dish catalog of the S2CLS UDS field \citep{geach2017scuba}, and the earlier ALESS survey \citep{karim2013alma}. To convert the S2CLS 850-$\mu$m counts to a common $S_{\rm 870}$ we use a factor of $S_{\rm 870}/S_{\rm 850}=$\,0.95 derived from a redshifted ($z=$\,2.5), composite spectral energy distribution (SED) for SMGs from the ALESS survey \citep{swinbank2013alma}, although we note that this correction is smaller than the estimated absolute calibration precision from S2CLS of 15\,\% \citep{geach2017scuba}.

%
%

Compared to a single power-law fit, the number counts of SMGs show a steepening decline at brighter fluxes. As a result the best fit to the differential number counts is with a double power-law function with the form:

\begin{equation}
   \frac{dN}{dS} = \frac{N_{0}}{S_{0}}\Big[\Big(\frac{S}{S_{0}}\Big)^{\alpha}+\Big(\frac{S}{S_{0}}\Big)^{\beta}\Big]^{-1},
	\label{eq:doublepower}
\end{equation}

where $N_{0}$ describes the normalisation, $S_{0}$ the break flux density, $\alpha$ and $\beta$ the two power-law slopes. For our AS2UDS data the best-fit parameters found are $N_{0}=1200^{+200}_{-300}$\,deg$^{-2}$, $S_{0}=5.1\pm0.7$\,mJy, $\alpha=5.9^{+1.3}_{-0.9}$ and $\beta=0.4\pm0.1$.

\begin{deluxetable}{ccc}[h]
\tablecaption{AS2UDS number counts \label{tab:countstab}}
\tablehead{
\colhead{$S_{\rm 870}$} & \colhead{$N(>S^{\prime}_{\rm 870})$\tablenotemark{a}} & \colhead{$dN/dS$\tablenotemark{b}}\\
\colhead{(mJy)} & \colhead{(deg$^{-2}$)}& \colhead{(mJy$^{-1}$\,deg$^{-2}$)}}
\startdata
4.5 & 385.3$^{+21.1}_{-7.7}$& 168.5$^{+14.8}_{-7.9}$\\
5.5 & 216.7$^{+17.3}_{-6.6}$ & 110.5$^{+12.1}_{-4.1}$\\
6.5 & 106.2$^{+11.4}_{-3.5}$ & 52.6$^{+8.3}_{-2.6}$\\
7.5 & 53.6$^{+8.4}_{-2.5}$ & 24.1$^{+6.0}_{-1.9}$\\
8.5 & 29.6$^{+6.5}_{-1.9}$ & 9.5$^{+4.2}_{-1.1}$\\
9.5 & 20.0$^{+5.7}_{-1.8}$ & 9.4$^{+4.2}_{-1.1}$\\
10.5 & 10.5$^{+4.4}_{-1.2}$ & 5.2$^{+3.5}_{-0.9}$\\
11.5 & 5.3$^{+3.5}_{-0.9}$ & 3.1$^{+3.0}_{-0.7}$\\
12.5 & 2.1$^{+2.8}_{-0.6}$ & --\\
13.5 & 2.1$^{+2.8}_{-0.6}$ & 1.0$^{+2.4}_{-0.5}$\\
14.5 & 1.0$^{+2.4}_{-0.5}$ & --\\
\enddata
\tablenotetext{a}{$S^{\prime}_{\rm 870}=S_{\rm 870}-0.5\Delta S$ where $\Delta S$ is 1\,mJy}
\tablenotetext{b}{``--'' denotes fluxes where there is no change in the cumulative counts between the lower flux bin and the current bin}
\end{deluxetable}

At $S_{\rm 870}\geq$\,4\,mJy we derive a surface density of 390$^{+70}_{-80}$\, deg$^{-2}$, corresponding to one  SMG per $\sim$\,arcmin$^{2}$ or one source per $\sim$\,130 ALMA primary beams at this frequency. Figure~\ref{fig:Counts} shows a systematic reduction in the surface density of SMGs compared to the single-dish estimate at all fluxes. This reduction from the SCUBA-2 counts to AS2UDS is statistically significant for sources fainter than $S_{\rm 870}=$\,8\,mJy, with a reduction of a factor of 28\,$\pm$\,2\,\% at $S_{\rm 870}\geq$\,4\,mJy and 41\,$\pm$\,8\,\% at $S_{\rm 870}\geq$\,7\,mJy. At the very bright end ($S_{\rm 870}\geq$\,12\,mJy) the number of SMGs is so low (just two in our $\sim$\,1\,deg$^2$ field) that the reduction in the relative number counts is poorly constrained, 30\,$\pm$\,20\,\%. Our bright-end reduction does agree with that seen in \cite{hill2018high} where they found a 24\,$\pm$\,6\,\% reduction between 11--15\,mJy in their SMA follow-up counts compared to the original SCUBA-2 parent sample. This agreement is unsurprising as a large number of their sources are drawn from our ALMA survey of the UDS field.    We also note that, as with our earlier pilot study of UDS in \citet{simpson2015scuba2}, that  we do not see an extreme drop-off of the counts above $S_{\rm 870}\sim$\,9\,mJy as was suggested from the smaller-area ALESS survey \citep{karim2013alma}.  

As we discuss below, the main factor which appears to be driving the 
the systematically lower counts of SMGs from interferometric studies, compared to the single-dish surveys,  
is that a fraction of the brighter single-dish sources break up into multiple fainter sources  (with flux densities of $S_{\rm 870}\lesssim$\,1--4\,mJy) in the interferometer maps and thus fall below the single-dish limit adopted for our counts. This effect has been termed ``multiplicity'' \cite{karim2013alma,simpson2015scuba2}.  An additional factor is the twelve ALMA ``blank'' maps of S2CLS sources brighter than $S^{\rm deb}_{\rm 870}\geq$\,4\,mJy, which also contribute to lowering the normalization of the number counts. These S2CLS sources, have a mean SNR  of 5.8\,$\pm$\,0.8, and are therefore unlikely to be spurious SCUBA-2 detections and our \textit{Herschel}/SPIRE stacking confirms this; instead the most likely explanation for their ALMA non-detection is ``extreme'' multiplicity, where the single-dish source breaks up into several faint SMGs below the detection limit of our ALMA maps. For these brighter SCUBA-2 sources with   ``blank'' ALMA maps this would require that the single-dish source  breaks up into $\geq$\,4 sources to result in a non-detection.

\subsection{Multiplicity}

There are differing claims in the literature regarding the influence of multiplicity of SMGs on single-dish submillimeter surveys. This is a result of both the differing  depths of the interferometric studies used to investigate this issue and the different definitions of ``multiplicity'' adopted in these works. Our survey has a relatively uniform sensitivity of $\sigma_{\rm 870}\sim$\,0.25\,mJy\,beam$^{-1}$, and therefore we adopt a fixed $S_{\rm 870}$ limit to identify multiple SMGs. We follow \cite{simpson2015scuba2} and define a multiple map as any field with more than one $S_{\rm 870}\geq$\,1\,mJy SMG within our ALMA Band 7 primary beam (i.e.\ within $\sim$\,9$\arcsec$ of the original SCUBA-2 detection locations). At the redshift of SMGs this corresponds to borderline U/LIRG systems, $L_{\rm IR}\geq$\,10$^{12}$\,$L_{\odot}$ which have SFRs of the order of 10$^{2}$\,M$_{\odot}$\,yr$^{-1}$ \citep{swinbank2013alma}. We also believe this is a more physical choice than, for example, using the {\it relative} submillimeter brightness  of the two sources to decide if they constitute a ``multiple'', as the relative fluxes may have little relevance to their other physical properties (e.g., mass or redshift) which are essential to understand their significance.

In our full sample we have maps with more than one  $S_{\rm 870}\geq1$\,mJy SMG in 79 of the 716 observations (11\,$\pm$\,1\,\%). We note that at 1\,mJy our ALMA observations are not complete, therefore this sets the multiplicity as a lower limit.
The surface density of  $S_{\rm 870}\sim$\,1\,mJy SMGs is $\sim$\,1\,arcmin$^{-2}$, as estimated from  unbiased ALMA surveys  \citep{aravena2016alma,dunlop2016deep}. Hence we expect to find one $S_{\rm 870}\sim$\,1\,mJy SMG per $\sim$\,19 ALMA primary beams or in $\sim$\,5\% of the maps, compared to the observed rate of $\sim$\,11\% (one per nine ALMA maps). We note, however, that the presence of a secondary source in these maps may act to increase the likelihood of the inclusion of that map into our sample by boosting the apparent SCUBA-2 flux into the S2CLS catalog. To address this potential bias we estimate the multiplicity rate for the 179 brighter single-dish sources with deboosted SCUBA-2 flux densities of $S^{\rm deb}_{\rm 850}\geq$\,5\,mJy.   The rate of multiples in these brighter SCUBA-2 sources is much higher 51/179 (28\,$\pm$\,2\,\%), suggesting that the presence of a {\it detected} secondary SMG in faint single-dish sources does not  strongly influence the inclusion of that single-dish source into our parent catalog.  Instead, the influence of multiplicity in faint single-dish sources is more likely to be seen through the presence of ``blank'' maps.  Hence we also place an upper limit on the multiplicity in our full survey by assuming that {\it all} the blank fields are a result of the blending of multiple faint SMGs, giving 187/716 (26\,$\pm$\,2\,\%) multiples. 

\begin{figure}[ht!]
\plotone{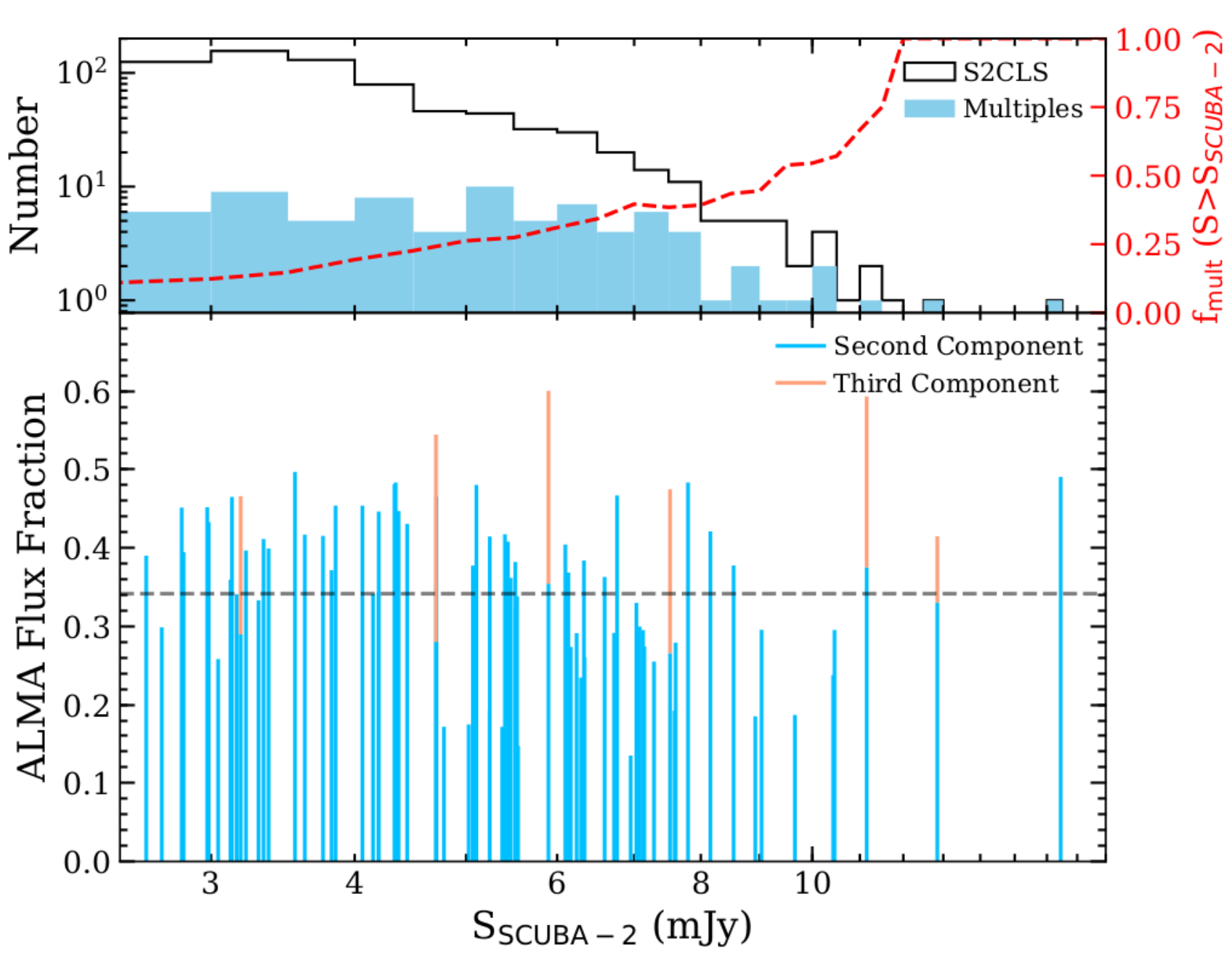}
\caption{\textit{Lower:} The fraction of the integrated ALMA flux of SMGs in each AS2UDS ALMA map that is contributed by secondary and tertiary components (ranked in terms of their relative brightness) as a function of the deboosted flux of the corresponding SCUBA-2 source. The horizontal dashed line shows the median fraction of the total flux contributed by secondary SMGs for these maps, 34\,$\pm$\,2\,\%. There is no significant trend in the  fractional flux density contributed by the secondary component as a function of the original SCUBA-2 flux density.
\textit{Upper:} The filled histogram show the distribution of the deboosted 850-$\mu$m fluxes of those SCUBA-2 sources that have multiple SMGs in our ALMA follow-up maps, and the unfilled histogram shows the corresponding SCUBA-2 fluxes of the parent sample of all 716 single-dish sources. We also plot cumulative fraction of the single-dish sources with fluxes greater than $S_{\rm{SCUBA-2}}$ that break up into multiple components, $f_{\rm{mult}} (S>S_{\rm{SCUBA-2}})$.  This fraction increases with increasing single-dish flux. \label{fig:Multiplicity}
}
\end{figure}

As implied above, the multiplicity appears to depend on the single-dish flux: as expected as the inclusion of emission from other SMGs
within the beam can only act to increase the apparent flux of the (blended) single-dish source.  
As described in \S 1, early observations suggested that roughly a third of  $S_{\rm 850}>$\,5\,mJy single-dish sources could be blends of multiple SMGs, with this rate increasing to 90\,\% for $S_{\rm 870}>$\,9\,mJy \citep[e.g.][]{karim2013alma}. As shown in Figure\,\,\ref{fig:Multiplicity}, for AS2UDS we find a frequency of multiplicity (ignoring ``blank'' maps) of 28\,$\pm$\,2\,\% for $S^{\rm deb}_{\rm 850}\geq$\,5\,mJy rising to 44$^{+16}_{-14}$\,\% at $S^{\rm deb}_{\rm 850}\geq$\,9\,mJy.

In Figure\,\ref{fig:Multiplicity} we also plot the fractional contribution of each secondary and tertiary ALMA SMG (ranked by flux density) to the total recovered ALMA flux density of all the SMGs for each field with multiple SMGs. The mean fraction of the total flux contributed by the secondary component  is 34\,$\pm$\,2\,\% with no significant variation of this fraction as a function of the original deboosted SCUBA-2 source flux. The 64\,$\pm$\,2\% contribution from the primary components in maps with multiple SMGs is broadly consistent with the semi-analytic model of \cite{cowley2014simulated} which suggested that $\sim$\,70\,\% of the flux density in blended sources would arise from the brightest component.  

%
%

\subsubsection{Physical association of the multiple SMGs}

Based on our Cycle 1 pilot study, \cite{simpson2015scuba2} showed that the number density of secondary SMGs in the maps of their 30 bright SCUBA-2 sources was 80\,$\pm$\,30 times that expected from blank-field number counts, suggesting that at least a fraction of these SMGs must be physically associated. Using our large sample we now seek to test this further. The most reliable route to test for physical association between SMGs in the same ALMA map would be to use spectroscopic redshifts for the  SMGs. However, as the current spectroscopic coverage of SMGs in AS2UDS is sparse, we instead exploit photometric redshifts to undertake this test.  We use the photometric redshift
 catalog constructed from the UKIDSS DR11 release (Hartley et al.\ in prep.), where a full description of the  DR11 observations will
be given in Almaini et al.\ (in prep.).  These photometric redshifts are derived from twelve photometric bands  ($U,B,V,R,I,z,Y,J,H,K,[3.6],[4.5]$) and applied to 296,007 $K$-band-detected sources using {\sc eazy} \citep{brammer2008eazy}; details of the methodology can be found in \cite{simpson2013prevalence}. The accuracy of these photometric redshifts is investigated in Hartley et al.\ (in prep.) from comparison with the $\sim$\,6,500 sources in the UKIDSS DR11 catalog which have spectroscopic redshifts, finding $|z_{\rm spec}-z_{\rm phot}|/(1+z_{\rm spec})=0.019\pm0.001$ with a median precision of $\sim$\,9\,\%. Around 85\,\% of the ALMA maps fall in regions of the UDS with high-quality photometric redshifts and these are considered in the following analysis.

\begin{figure}[ht!]
\plotone{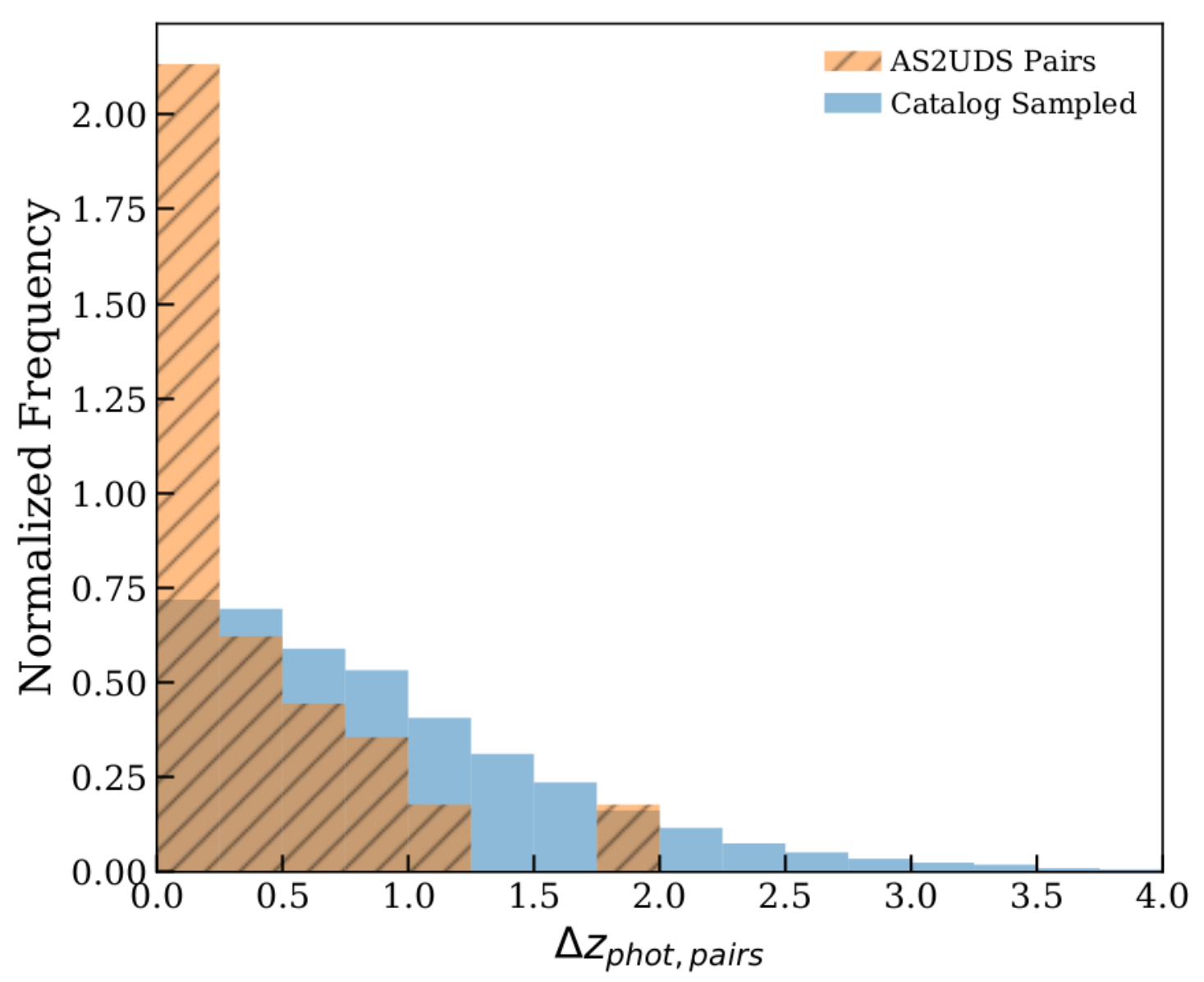}
\caption{The normalized distribution of redshift separation, $\Delta z_{\rm phot}$, for pairs of SMGs with reliable photometric redshifts detected in the same ALMA map (separation $\lesssim$\,9\arcsec), compared to pairs of SMGs randomly selected from the distribution of all isolated AS2UDS SMGs with photometric redshifts. The strong peak at $\Delta z_{\rm phot}<$\,0.25 for the SMGs pairs compared to the random sample, which occurs less than 1\,\% of the time by chance in our simulations, suggests that a moderate fraction of multiple SMGs (at least those with optically bright counterparts) in single fields arise from physically associated galaxies, rather than chance line of sight projections.   \label{fig:Deltaz}}
\end{figure}

In Figure\,\ref{fig:Deltaz} we plot the distribution of the differences in photometric redshifts ($\Delta z_{\rm phot}$) for pairs of SMGs in those single-dish maps with multiple ALMA-detected SMGs. We limit our analysis to SMGs that  fall within the region with high-quality photometric redshifts and which have $K$-band detections within 0$\farcs$6 radius from the ALMA  positions (497 of the 695 SMGs) for both sources in the map.   This yields 46 pairs of SMGs (92 SMGs in total) from the 164 SMGs in the 79 maps with multiple SMGs.  We find that 52\,\% of these pairs (24/46) have $\Delta z_{\rm phot} <$\,0.25. We note that 2$\arcsec$ diameter apertures were employed for the photometry in the DR11 catalog, therefore the $\Delta z_{\rm phot}$ was additionally calculated for only pairs that are separated by greater than 2$\arcsec$, thus removing the possibility of neighbours contaminating photometry and thus photometric redshifts. This \textit{still} results in 53\,\% of pairs having $\Delta z_{\rm phot} <$\,0.25 (23/43). 

To assess the significance of this result we next quantify whether the 24 pairs of blended SMGs with $\Delta z_{\rm phot}<0.25$ is statistically in excess of expectations for 46 random SMG pairs. To do this we determine the expected distribution of $\Delta z_{\rm phot}$ for pairs of SMGs randomly selected from the 497 SMGs with high-quality photometric redshifts across the full field, and plot this in Figure\,\ref{fig:Deltaz}.  To perform this test we sample the random distribution of our unassociated SMGs 10,000 times, each time drawing 46 pairs, and testing how frequently $>$\,52\,\% of these are found to have  $\Delta z_{\rm phot} <$\,0.25. This analysis shows that the median fraction of random pairs with $\Delta z_{\rm phot} <$\,0.25 is 20\,$\pm$\,2\,\% compared to the 52\,\% for the actual pairs of SMGs. This strongly suggests that a significant fraction of the single-dish sources that resolve into multiple optically-bright (e.g.\ those with photometric redshifts) SMGs are in fact physically associated galaxies on projected angular scales of $\sim$\,10--100\,kpc scales. If we assume that all pairs without photometric redshifts for \textit{both} SMGs are physically unassociated, a conservative estimate, then comparing to the total number of ALMA fields with multiple SMGs, we can place a lower limit of at least 30\,\% (24 pairs out of 79) on the fraction of all multiple-SMG fields arising from closely associated galaxies. This is consistent with previous spectroscopic studies of SMG multiples e.g.\ $\sim$\,40\,\% of SMG pairs  physically associated combining the estimates from \cite{wardlow2018} and \cite{hayward2018observational}. Of course, to truly test this requires a spectroscopic redshift survey of a much large sample of these multiple-SMG systems. 

%
%

%
%
%
\section{Conclusions} \label{sec:conclusion}
We have presented the first results from a large ALMA 870-$\mu$m continuum survey of  716 single-dish submillimeter sources drawn from the SCUBA-2 Cosmology Legacy Survey map of the UKIDSS UDS field. These sensitive, high-resolution ALMA observations provide the largest sample of interferometrically detected submillimeter galaxies constructed to date, with 695 SMGs above 4.3\,$\sigma$ (corresponding to a false detection rate of 2\,\%). This sample is seven times larger in terms number of SMGs and drawn from a single-dish survey which has four times the area of the previous largest interferometric SMG survey. The main conclusions of this work are as follows:

\begin{itemize}
\item We construct resolved 870\,$\mu$m differential and cumulative number counts brighter than S$_{\rm 870}\geq$\,4\,mJy (a conservative choice based on the flux limit of the parent single-dish S2CLS survey), which show a similar shape to the number counts from S2CLS, but with a systematically lower normalization at fixed flux density,  by a factor of 1.28$\pm$0.02. Much of this reduction in the SMG counts, is due to the influence of multiplicity, i.e.\ single-dish sources splitting into two or more SMGs detected by ALMA.  We fit a double power-law function to our differential number counts to easily facilitate future comparison with observations in other fields and simulations.

\item In 11\,$\pm$\,1\,\% of our 716 ALMA maps we detect more than one SMG with S$_{\rm 870}\geq$\,1\,mJy corresponding to a $L_{\rm IR} \geq $\,10$^{12}$ L$_{\odot}$ galaxy in a region with a projected diameter of $\sim$\,100\,kpc at $z=$\,2.  This multiplicity fraction varies from 26\,$\pm$\,2\,\% for all single-dish sources with $S^{\rm deb}_{\rm 850}\geq$\,5\,mJy, to 44$^{+16}_{-14}$\,\% at $S^{\rm deb}_{\rm 850}\geq$\,9\,mJy. The brightest of these multiple-SMG components typically contributes 64\,$\pm$\,2\,\% of the total flux of the SCUBA-2 source, with no detectable variation in this fraction with with single-dish source flux, consistent with results from semi-analytic models of blending in single-dish surveys.

\item By comparing the photometric redshift differences between pairs of SMGs in ALMA maps with multiple components, we show evidence that a significant fraction of these pairs are  likely to be physically associated, with $\gtrsim$\,30\,\% of all multiple-SMG maps arising from physically associated galaxies.
\end{itemize}

\acknowledgments

SMS\ acknowledges the support of STFC studentship (ST/N50404X/1). AMS\ and IS\ acknowledge financial support from an STFC grant (ST/P000541/1). IS\, EAC\, and BG\ also acknowledge support from the ERC Advanced Investigator program DUSTYGAL 321334, and a Royal Society/Wolfson Merit Award. JEG acknowledges support from a Royal Society University Research Fellowship. JLW acknowledges the support of an STFC Ernest Rutherford Fellowship. MJM acknowledges the support of the National Science Centre, Poland through the POLONEZ grant 2015/19/P/ST9/04010; this project has received funding from the European Union's Horizon 2020 research and innovation programme under the Marie Sk{\l}odowska-Curie grant agreement No. 665778. The ALMA data used in this paper were obtained under programs ADS/JAO.ALMA\#2012.1.00090.S, \#2015.1.01528.S and \#2016.1.00434.S. ALMA is a partnership of ESO (representing its member states), NSF (USA) and NINS (Japan), together with NRC (Canada) and NSC and ASIAA (Taiwan), in cooperation with the Republic of Chile. The Joint ALMA Observatory is operated by ESO, AUI/NRAO, and NAOJ.  This paper used data from project MJLSC02 on the James Clerk Maxwell Telescope, which is operated by the East Asian Observatory on behalf of The National Astronomical Observatory of Japan, Academia Sinica Institute of Astronomy and Astrophysics, the Korea Astronomy and Space Science Institute, the National Astronomical Observatories of China and the Chinese Academy of Sciences (Grant No.\ XDB09000000), with additional funding support from the Science and Technology Facilities Council of the United Kingdom and participating universities in the United Kingdom and Canada. UKIDSS-DR11 photometry made use of UKIRT. UKIRT is owned by the University of Hawaii (UH) and operated by the UH Institute for Astronomy; operations are enabled through the cooperation of the East Asian Observatory. When (some of) the data reported here were acquired, UKIRT was supported by NASA and operated under an agreement among the University of Hawaii, the University of Arizona, and Lockheed Martin Advanced Technology Center; operations were enabled through the cooperation of the East Asian Observatory. When (some of) the data reported here were acquired, UKIRT was operated by the Joint Astronomy Centre on behalf of the Science and Technology Facilities Council of the U.K.

\bibliographystyle{aasjournal}
\bibliography{Countsref} 

\begin{thebibliography}{}
\expandafter\ifx\csname natexlab\endcsname\relax\def\natexlab#1{#1}\fi
\providecommand{\url}[1]{\href{#1}{#1}}

\bibitem[{Aravena {et~al.}(2016)Aravena, Decarli, Walter, Da~Cunha, Bauer,
  Carilli, Daddi, Elbaz, Ivison, Riechers, {et~al.}}]{aravena2016alma}
Aravena, M., Decarli, R., Walter, F., {et~al.} 2016, ApJ, 833, 68

\bibitem[{Barger {et~al.}(1998)Barger, Cowie, Sanders, Fulton, Taniguchi, Sato,
  Kawara, \& Okuda}]{barger1998submillimetre}
Barger, A., Cowie, L., Sanders, D., {et~al.} 1998, Nature, 394, 248

\bibitem[{Blain {et~al.}(2002)Blain, Smail, Ivison, Kneib, \&
  Frayer}]{blain2002submillimeter}
Blain, A.~W., Smail, I., Ivison, R., Kneib, J.-P., \& Frayer, D.~T. 2002, Phys.
  Reports, 369, 111

\bibitem[{Brammer {et~al.}(2008)Brammer, van Dokkum, \&
  Coppi}]{brammer2008eazy}
Brammer, G.~B., van Dokkum, P.~G., \& Coppi, P. 2008, ApJ, 686, 1503

\bibitem[{Casey {et~al.}(2013)Casey, Chen, Cowie, Barger, Capak, Ilbert, Koss,
  Lee, Le~Floc'h, Sanders, {et~al.}}]{casey2013characterization}
Casey, C.~M., Chen, C.-C., Cowie, L.~L., {et~al.} 2013, MNRAS, 436, 1919

\bibitem[{Chapman {et~al.}(2005)Chapman, Blain, Smail, \&
  Ivison}]{chapman2005redshift}
Chapman, S.~C., Blain, A., Smail, I., \& Ivison, R. 2005, ApJ, 622, 772

\bibitem[{Chen {et~al.}(2013)Chen, Cowie, Barger, Casey, Lee, Sanders, Wang, \&
  Williams}]{chen2013resolving}
Chen, C.-C., Cowie, L.~L., Barger, A.~J., {et~al.} 2013, ApJ, 776, 131

\bibitem[{Chen {et~al.}(2016)Chen, Smail, Ivison, Arumugam, Almaini, Conselice,
  Geach, Hartley, Ma, Mortlock, {et~al.}}]{chen2016scuba}
Chen, C.-C., Smail, I., Ivison, R.~J., {et~al.} 2016, ApJ, 820, 82

\bibitem[{Cimatti {et~al.}(2008)Cimatti, Cassata, Pozzetti, Kurk, Mignoli,
  Renzini, Daddi, Bolzonella, Brusa, Rodighiero, {et~al.}}]{cimatti2008gmass}
Cimatti, A., Cassata, P., Pozzetti, L., {et~al.} 2008, A\&A, 482, 21

\bibitem[{Condon(1997)}]{condon1997errors}
Condon, J. 1997, PASP, 109, 166

\bibitem[{Condon {et~al.}(1998)Condon, Cotton, Greisen, Yin, Perley, Taylor, \&
  Broderick}]{condon1998nrao}
Condon, J.~J., Cotton, W., Greisen, E., {et~al.} 1998, ApJ, 115, 1693

\bibitem[{Coppin {et~al.}(2008)Coppin, Swinbank, Neri, Cox, Alexander, Smail,
  Page, Stevens, Knudsen, Ivison, {et~al.}}]{coppin2008testing}
Coppin, K., Swinbank, A., Neri, R., {et~al.} 2008, MNRAS, 389, 45

\bibitem[{Cowley {et~al.}(2015)Cowley, Lacey, Baugh, \&
  Cole}]{cowley2014simulated}
Cowley, W.~I., Lacey, C.~G., Baugh, C.~M., \& Cole, S. 2015, MNRAS, 446, 1784

\bibitem[{Dunlop {et~al.}(2016)Dunlop, McLure, Biggs, Geach, Micha{\l}owski,
  Ivison, Rujopakarn, van Kampen, Kirkpatrick, Pope, {et~al.}}]{dunlop2016deep}
Dunlop, J.~S., McLure, R., Biggs, A., {et~al.} 2016, MNRAS, 466, 861

\bibitem[{Farrah {et~al.}(2006)Farrah, Lonsdale, Borys, Fang, Waddington,
  Oliver, Rowan-Robinson, Babbedge, Shupe, Polletta,
  {et~al.}}]{farrah2006spatial}
Farrah, D., Lonsdale, C., Borys, C., {et~al.} 2006, ApJL, 641, L17

\bibitem[{Franco {et~al.}(2018)Franco, Elbaz, B{\'e}thermin, Magnelli,
  Schreiber, Ciesla, Dickinson, Nagar, Silverman, Daddi,
  {et~al.}}]{franco2018goods}
Franco, M., Elbaz, D., B{\'e}thermin, M., {et~al.} 2018, arXiv preprint
  arXiv:1803.00157

\bibitem[{Fu {et~al.}(2013)Fu, Cooray, Feruglio, Ivison, Riechers, Gurwell,
  Bussmann, Harris, Altieri, Aussel, {et~al.}}]{fu2013rapid}
Fu, H., Cooray, A., Feruglio, C., {et~al.} 2013, Nature, 498, 338

\bibitem[{Geach {et~al.}(2017)Geach, Dunlop, Halpern, Smail, van~der Werf,
  Alexander, Almaini, Aretxaga, Arumugam, Asboth, {et~al.}}]{geach2017scuba}
Geach, J., Dunlop, J., Halpern, M., {et~al.} 2017, MNRAS, 465, 1789

\bibitem[{Hayward {et~al.}(2018)Hayward, Chapman, Steidel, Golob, Casey, Smith,
  Zitrin, Blain, Bremer, Chen, {et~al.}}]{hayward2018observational}
Hayward, C.~C., Chapman, S.~C., Steidel, C.~C., {et~al.} 2018, MNRAS, 476, 2278

\bibitem[{Hickox {et~al.}(2012)Hickox, Wardlow, Smail, Myers, Alexander,
  Swinbank, Danielson, Stott, Chapman, Coppin, {et~al.}}]{hickox2012laboca}
Hickox, R.~C., Wardlow, J., Smail, I., {et~al.} 2012, MNRAS, 421, 284

\bibitem[{Hill {et~al.}(2018)Hill, Chapman, Scott, Petitpas, Smail, Chapin,
  Gurwell, Perry, Blain, Bremer, {et~al.}}]{hill2018high}
Hill, R., Chapman, S.~C., Scott, D., {et~al.} 2018, MNRAS, 477, 2042

\bibitem[{Hodge {et~al.}(2013)Hodge, Karim, Smail, Swinbank, Walter, Biggs,
  Ivison, Weiss, Alexander, Bertoldi, {et~al.}}]{hodge2013alma}
Hodge, J., Karim, A., Smail, I., {et~al.} 2013, ApJ, 768, 91

\bibitem[{Hughes {et~al.}(1998)Hughes, Serjeant, Dunlop, Rowan-Robinson, Blain,
  Mann, Ivison, Peacock, Efstathiou, Gear, {et~al.}}]{hughes1998high}
Hughes, D.~H., Serjeant, S., Dunlop, J., {et~al.} 1998, Nature, 394, 241

\bibitem[{Ikarashi {et~al.}(2011)Ikarashi, Kohno, Aguirre, Aretxaga, Arumugam,
  Austermann, Bock, Bradford, Cirasuolo, Earle,
  {et~al.}}]{ikarashi2011detection}
Ikarashi, S., Kohno, K., Aguirre, J., {et~al.} 2011, MNRAS, 415, 3081

\bibitem[{Ivison {et~al.}(2007)Ivison, Greve, Dunlop, Peacock, Egami, Smail,
  Ibar, Van~Kampen, Aretxaga, Babbedge, {et~al.}}]{ivison2007scuba}
Ivison, R.~J., Greve, T., Dunlop, J., {et~al.} 2007, MNRAS, 380, 199

\bibitem[{Karim {et~al.}(2013)Karim, Swinbank, Hodge, Smail, Walter, Biggs,
  Simpson, Danielson, Alexander, Bertoldi, {et~al.}}]{karim2013alma}
Karim, A., Swinbank, A., Hodge, J., {et~al.} 2013, MNRAS, stt196

\bibitem[{Lacey {et~al.}(2016)Lacey, Baugh, Frenk, Benson, Bower, Cole,
  Gonzalez-Perez, Helly, Lagos, \& Mitchell}]{lacey2016unified}
Lacey, C.~G., Baugh, C.~M., Frenk, C.~S., {et~al.} 2016, MNRAS, 462, 3854

\bibitem[{Lawrence {et~al.}(2007)Lawrence, Warren, Almaini, Edge, Hambly,
  Jameson, Lucas, Casali, Adamson, Dye, {et~al.}}]{lawrence2007ukirt}
Lawrence, A., Warren, S., Almaini, O., {et~al.} 2007, MNRAS, 379, 1599

\bibitem[{Lilly {et~al.}(1999)Lilly, Eales, Gear, Hammer, Le~Fevre, Crampton,
  Bond, \& Dunne}]{lilly1999canada}
Lilly, S.~J., Eales, S.~A., Gear, W.~K., {et~al.} 1999, ApJ, 518, 641

\bibitem[{Magnelli {et~al.}(2012)Magnelli, Lutz, Santini, Saintonge, Berta,
  Albrecht, Altieri, Andreani, Aussel, Bertoldi,
  {et~al.}}]{magnelli2012herschel}
Magnelli, B., Lutz, D., Santini, P., {et~al.} 2012, A\&A, 539, A155

\bibitem[{McMullin {et~al.}(2007)McMullin, Waters, Schiebel, Young, \&
  Golap}]{mcmullin2007casa}
McMullin, J., Waters, B., Schiebel, D., Young, W., \& Golap, K. 2007, in
  Astronomical data analysis software and systems XVI, Vol. 376, 127

\bibitem[{Simpson {et~al.}(2013)Simpson, Westoby, Arumugam, Ivison, Hartley, \&
  Almaini}]{simpson2013prevalence}
Simpson, C., Westoby, P., Arumugam, V., {et~al.} 2013, MNRAS, 433, 2647

\bibitem[{Simpson {et~al.}(2015{\natexlab{a}})Simpson, Smail, Swinbank,
  Chapman, Geach, Ivison, Thomson, Aretxaga, Blain, Cowley,
  {et~al.}}]{simpson2015scuba2}
Simpson, J., Smail, I., Swinbank, A., {et~al.} 2015{\natexlab{a}}, ApJ, 807,
  128

\bibitem[{Simpson {et~al.}(2015{\natexlab{b}})Simpson, Smail, Swinbank,
  Almaini, Blain, Bremer, Chapman, Chen, Conselice, Coppin,
  {et~al.}}]{simpson2015scuba}
---. 2015{\natexlab{b}}, ApJ, 799, 81

\bibitem[{Simpson {et~al.}(2017)Simpson, Smail, Swinbank, Ivison, Dunlop,
  Geach, Almaini, Arumugam, Bremer, Chen, {et~al.}}]{simpson2017scuba}
---. 2017, ApJ, 839, 58

\bibitem[{Simpson {et~al.}(2014)Simpson, Swinbank, Smail, Alexander, Brandt,
  Bertoldi, de~Breuck, Chapman, Coppin, da~Cunha, {et~al.}}]{simpson2014alma}
Simpson, J.~M., Swinbank, A., Smail, I., {et~al.} 2014, ApJ, 788, 125

\bibitem[{Smail {et~al.}(1997)Smail, Ivison, \& Blain}]{smail1997deep}
Smail, I., Ivison, R., \& Blain, A. 1997, ApJL, 490, L5

\bibitem[{Swinbank {et~al.}(2013)Swinbank, Simpson, Smail, Harrison, Hodge,
  Karim, Walter, Alexander, Brandt, de~Breuck, {et~al.}}]{swinbank2013alma}
Swinbank, A., Simpson, J., Smail, I., {et~al.} 2013, MNRAS, 438, 1267

\bibitem[{Tacconi {et~al.}(2006)Tacconi, Neri, Chapman, Genzel, Smail, Ivison,
  Bertoldi, Blain, Cox, Greve, {et~al.}}]{tacconi2006high}
Tacconi, L.~J., Neri, R., Chapman, S., {et~al.} 2006, ApJ, 640, 228

\bibitem[{Toft {et~al.}(2014)Toft, Smol{\v{c}}i{\'c}, Magnelli, Karim, Zirm,
  Michalowski, Capak, Sheth, Schawinski, Krogager,
  {et~al.}}]{toft2014submillimeter}
Toft, S., Smol{\v{c}}i{\'c}, V., Magnelli, B., {et~al.} 2014, ApJ, 782, 68

\bibitem[{Wang {et~al.}(2010)Wang, Cowie, Barger, \& Williams}]{wang2010sma}
Wang, W.-H., Cowie, L.~L., Barger, A.~J., \& Williams, J.~P. 2010, ApJL, 726,
  L18

\bibitem[{Wardlow {et~al.}(2011)Wardlow, Smail, Coppin, Alexander, Brandt,
  Danielson, Luo, Swinbank, Walter, Wei{\ss}, {et~al.}}]{wardlow2011laboca}
Wardlow, J., Smail, I., Coppin, K., {et~al.} 2011, MNRAS, 415, 1479

\bibitem[{Wardlow {et~al.}(2018)Wardlow, Smail, Swinbank,
  {et~al.}}]{wardlow2018}
Wardlow, J.~L., Smail, I., Swinbank, A., {et~al.} 2018, MNRAS, submitted

\bibitem[{Wei{\ss} {et~al.}(2009)Wei{\ss}, Kov{\'a}cs, Coppin, Greve, Walter,
  Smail, Dunlop, Knudsen, Alexander, Bertoldi, {et~al.}}]{weiss2009large}
Wei{\ss}, A., Kov{\'a}cs, A., Coppin, K., {et~al.} 2009, ApJ, 707, 1201

\bibitem[{Whitaker {et~al.}(2012)Whitaker, Kriek, Van~Dokkum, Bezanson,
  Brammer, Franx, \& Labb{\'e}}]{whitaker2012large}
Whitaker, K.~E., Kriek, M., Van~Dokkum, P.~G., {et~al.} 2012, ApJ, 745, 179

\bibitem[{Wilkinson {et~al.}(2016)Wilkinson, Almaini, Chen, Smail, Arumugam,
  Blain, Chapin, Chapman, Conselice, Cowley, {et~al.}}]{wilkinson2016scuba}
Wilkinson, A., Almaini, O., Chen, C.-C., {et~al.} 2016, MNRAS, 464, 1380

\bibitem[{Zavala {et~al.}(2015)Zavala, Yun, Aretxaga, Hughes, Wilson, Geach,
  Egami, Gurwell, Wilner, Smail, {et~al.}}]{zavala2015early}
Zavala, J., Yun, M., Aretxaga, I., {et~al.} 2015, MNRAS, 452, 1140

\end{thebibliography}


\end{document}